\documentclass[twocolumn]{revtex4-1}
\usepackage{graphicx}
\usepackage{amssymb}
\usepackage{amsfonts}
\usepackage{amsbsy}
\usepackage{dcolumn}
\usepackage{color}
\usepackage[english]{babel}
\usepackage{epsfig}
\usepackage{multirow}
\usepackage{natbib}
\begin{document}

\title{Probing bulk viscous matter-dominated models with Gamma-ray bursts}

\author{A. Montiel and N. Bret\'on}
\affiliation{Dpto. de F\'isica, Centro de Investigaci\'on y de Estudios Avanzados del I. P. N.,\\ Av. IPN 2508, D.F., M\'exico}

\begin{abstract}
In this paper we extend the range of consistency of a constant bulk viscosity model to redshifts up to  $z\sim 8.1$. In this model the dark sector of the cosmic substratum is a viscous fluid with pressure $p= -\zeta \theta$, where $\theta$ is the fluid-expansion  scalar and $\zeta$ is the coefficient of bulk viscosity. Using the sample of 59 high-redshift GRBs reported by Wei (2010), we calibrate GRBs at low redshifts with the Union 2 sample of SNe Ia, avoiding then the circularity problem. 
Testing the constant bulk viscosity model with GRBs we found the best fit for the viscosity parameter $\tilde{\zeta}$ in the range $0<\tilde{\zeta}<3$, being so consistent with previous probes; we also determined the deceleration parameter $q_0$ and the redshift of transition to accelerated expansion.
Besides we present an updated analysis of the model with CMB5-year data and CMB7-year data, as well as with the baryon acoustic peak BAO. From the statistics with CMB it turns out that the model does not describe in a feasible way the far far epoch of recombination of the universe, but is in very good concordance for epochs as far as $z\sim 8.1$ till present.
\end{abstract}
\maketitle

\section{Introduction}

Since 1998 reliable cosmological data have been accumulated leading to the conclusion that our universe has entered recently into an accelerated expansion epoch. The luminosity-distance data of supernovae Ia (SNe Ia) has provided the main evidence. Hitherto explanations to this phenomenon 
are not satisfactory. Most models consider accelerated expansion 
as due to a component of the universe
that behaves opposite to gravity, the so called \textit{dark energy} (DE). Another unknown component of the universe is the \textit{dark matter} (DM), the missing mass necessary to held together galaxy clusters, also needed to explain the current large scale structure of the universe.
The preferred model to describe accelerated expansion is the Lambda-Cold-Dark-Matter ($\Lambda$CDM).
However, cosmological constant has several theoretical drawbacks, like the discrepancies between the observed and theoretical values, i.e. the fine tuning problem; the coincidence problem is another objection, meaning the fact that dark energy is nearly equal to the matter density just now. Among many proposals to give a satisfactory answer there are some models that consider DE as a dynamical component, such as Quintessence or K-essence; in others proposals DE appears as the result of the interaction of fundamental particles. In here we adhere to the stream that considers DM and DE as different manifestations of the same component, describing the dark sector as
some kind of matter whose physical properties depend on the scale: behaving as DM at high densities and transforming into DE at lower densities. These Unified Dark Matter models have the generalized Chaplygin gas as its paradigmatic example \cite{Kamenshchik:2001, Bilic:2002, Bento:2002}, however these models present undesirable features like oscillations in the matter power spectrum and to mend this inconvenience it should be assumed the existence of nonadiabatic pressure perturbations.

On the other side, in the context of inflation of the very early Universe, it has been known since long time ago that an imperfect fluid with bulk viscosity can produce an acceleration in the expansion without falling back on a cosmological constant or some inflationary scalar field \cite{Heller:1973, Heller:1975, Zimdahl:1996}. Inspired in those inflationary models there are some recent \cite{Colistete:2007, Meng:2007, Ren:2007, Ren:2006, Ren:2006a, Meng:2009, Nucamendi:2009} and not so recent 
\cite{Padma:1987} developments that assume a universe filled with a viscous single fluid. The bulk viscosity contributes to the cosmic pressure and drives the accelerated expansion. 

The origin of the bulk viscosity in a physical system is due to its deviations from the local thermodynamic equilibrium
\cite{Israel:1963, Weinberg:1971} and different cooling rates \cite{Zimdahl:1996a, Zimdahl:1997}. In a cosmological fluid, the bulk viscosity arises when the fluid expands (or contracts) too fast so that the system does not have enough time to restore the local thermodynamic equilibrium and then it appears an effective pressure restoring the system to its local thermodynamic equilibrium. Bulk viscosity can be seen as a measurement of this effective pressure. When the fluid reaches again the thermal equilibrium then the bulk viscosity pressure ceases \cite{Wilson:2007}, \cite{Okumura:2003, Ilg:1999, Xinzhong:2001}.
So, in an accelerated expanding universe it may be natural to suppose that the expansion process is actually a collection of states out of thermal equilibrium in a small fraction of time that being modeled by a bulk viscous fluid might be a more realistic description of the accelerated universe today. 

Moreover, these models have in their favour that behave well when density perturbations come into play: the power spectrum for viscous matter is well behaved and consistent with large scale structure data \cite{Zimdahl:2010}. In particular, it does not suffer from the oscillation problem of Chaplygin models \cite{Zimdahl:2009}.
 
On the other hand, since the earliest evidence of tight correlations in gamma-ray bursts spectral properties \cite{Amati:2002}, the possibility arose of using GRBs as standard candles.
Being so GRBs may open a window in redshift as far as $z \sim 8$ \cite{Salvaterra}, extending then the attainable range provided by SNe Ia observations. For an overview of most recent missions, see \cite{McBreen} and references therein.
At these epochs, $z \sim 8$, the universe was dominated by dark matter, from which follows that this tool is less sensitive to dark energy. However for models where dark energy and matter are coupled \cite{Amendola:2003, Bertolami:2007} or unified \cite{Bento:2002} GRBs might be a useful tool. 

GRBs were prevented of being used as standard candles 
because the intrinsic faintness of the nearby events, a fact that introduced a bias towards low redshifts of GRB and therefore the extrapolation of their correlations to low-z events faced serious problems. 
However this problem is solved if one uses SNe Ia data, in a combined calibration with GRBs at low redshifts, allowing then GRBs be considered as distance indicators \cite{Ghirlanda, Amati:2008, Bertolami:2006, Liang:2006}. 
In particular, Liang \cite{Liang:2008} implemented a method to circumvent this objection, obtaining the distance modulus of GRB at low redshift by interpolating from the Hubble diagram of SNe Ia. Then GRB relations are calibrated without assumming a particular cosmological model; fair is to say that there is some criticism to the method \cite{Graziani}.

In this paper we test the model (presented in \cite{Nucamendi:2009}) that considers a bulk viscous fluid as source of matter, with GRBs data given in \cite{Wei:2010} containing redshifts up to $z=8.1$.
In this way we extend the previous SNe Ia probes of the model. In Sec. II we introduced the main aspects of the bulk viscosity model; in Sec. III the calibration of GRBs is addressed. In Sec. IV we apply the calibrated relation to higher redshifts and do the statistics to find the best fit for the bulk viscosity parameter. 
We carry on the analysis with the observational data including SNe Ia; BAO for clusters with redshifts up approximately 2, as well as CMB5-year and CMB7-year data from WMAP; this last probe tests the model up to the last scattering surface,  about $z\simeq1091.3$; according to our analysis the model is not reliable to such far epoch, but we extend its applicability to epochs of formation of structure $z \sim 8$. We discuss conclusions and perspectives in the last section.

\section{Bulk viscosity driving the accelerated expansion.}
 
In \cite{Nucamendi:2009}, the baryon and dark matter components are modeled by a pressureless fluid characterized by a constant bulk viscosity $\tilde{\zeta}$. Lying the bulk viscosity constant in the range $0<\tilde{\zeta}<3$, the model possesses many desirable properties, namely: holding the second law of thermodynamics, the derived age of the universe is in perfect agreement with the constraint of globular clusters, it has a Big-Bang followed by a decelerated expansion with a smooth transition to an accelerated expansion epoch in late times. Previous cosmological probes done by Avelino and Nucamendi include the SNe Ia Gold 2006 sample (182 SNe Ia) and BAO in a quick fit \cite{Avelino}. The many good properties of the model justifies to complete with GRBs the observational confrontation. In this paper the model is tested for the first time using GRBs, and updated tests using CMB five and seven years WMAP data, as well as SNe Ia Union2 sample and BAO. The results point out the strengths and what maybe be the limit of applicability of the model, as are the far far epochs of the universe, at very high redshifts of the recombination epoch.
 
Before going on, we note that Nucamendi-Avelino \cite{Nucamendi:2010} addressed the possibility of extending the model of constant bulk viscosity to a model where the viscosity depends on $H$ as $\zeta=\zeta_0+\zeta_1H$, where $\zeta_0$ and $\zeta_1$ are constants and $H$ is the Hubble parameter. Using the SCP Union2 SNe Ia data set composed of 557 type Ia SNe \cite{Union2}, they found that from all possible scenarios predicted by the model according to different values of the dimensionless bulk viscous coefficients $\tilde{\zeta}_0$ and $\tilde{\zeta}_1$ the preferred ones are two ($\tilde{\zeta}_0>0$, $\tilde{\zeta}_1<0$) with $\tilde{\zeta}_0+\tilde{\zeta}_1<3$, and ($0<\tilde{\zeta}_0<3$, $\tilde{\zeta}_1=0$). Nevertheless, one of the disadvantages of the model parameterized by $\tilde{\zeta}=\tilde{\zeta}_0+\tilde{\zeta}_1H$, is that the best estimated total bulk viscosity function $\tilde{\zeta}(z)$ is positive for redshifts $z \leq 1$ and negative for $z \geq 1$ latter implies a violation to the local second law of thermodynamics (see \cite{Nucamendi:2010} for details). Then, the simplest model of one parameter $\tilde{\zeta}=\tilde{\zeta}_0$ turned out to be the best candidate to explain the present accelerated expansion.

For an imperfect fluid, the energy-momentum tensor with a first-order deviation from the thermodynamic equilibrium is:
\begin{equation}
T_{\mu \nu}= \rho u_{\mu} u_{\nu} + (g_{\mu \nu}+u_{\mu} u_{\nu})P^*,
\label{eq:Tmu}
\end{equation}
with 
\begin{equation}
P^*\equiv P-\zeta\nabla^{\nu}u_{\nu},
\label{eq:P*}
\end{equation}
where $u_{\nu}$ is the four-velocity vector of an observer who measures the effective pressure $P^*$; $P$ and $\rho$ are the pressure and density of the fluid respectively. The term $\zeta$ is the bulk viscous coefficient that arises in the fluid which is out of local thermodynamic equilibrium. The conservation equation for the viscous fluid is
\begin{equation}
u^{\nu}\nabla_{\nu}\rho+(\rho+P^*)\nabla^{\nu}u_{\nu}=0.
\label{eq:conservation}
\end{equation} 

In \cite{Nucamendi:2009} it has been considered a pressureless fluid ($P=0$), then $P^*=-\zeta\nabla^{\nu}u_{\nu}$. We shall consider a spatially flat geometry for the Friedmann-Robertson-Walker (FRW) cosmology 
\begin{equation}
ds^2= -dt^2+a^2(t)(dr^2+r^2d\Omega^2),
\label{eq:metrica}
\end{equation}
where the function $a(t)$ is the scale factor. The conservation equation \ref{eq:conservation}
now becomes,

\begin{equation}
\dot{\rho}_m+(\rho_m-3H\zeta)3H=0,
\end{equation}
where $H\equiv \dot{a}/a$ is the Hubble parameter, $\rho_m$ is the total matter density (the dot means time derivative) and $\nabla^{\nu}u_{\nu}=3H$.
 
From the Friedmann equations, the Hubble parameter is given through
\begin{equation}
\rho_m-3H\zeta=\rho_m-H_0 \tilde{\zeta}\left({\frac{\rho_m}{24 \pi G}}\right)^{1/2}
\label{eq:Hz1}
\end{equation}
where $H_0$ is the Hubble constant today and it has been defined the dimensionless bulk viscous coefficient $\tilde{\zeta}$,
the matter density parameter $\Omega_{m0}$ and the critical density today $\rho^0_{\rm crit}$ as

\begin{equation}
\tilde{\zeta}\equiv \frac{24\pi G}{H_0}\zeta,\quad \Omega_{m0}\equiv \frac{\rho_{m0}}{\rho^0_{\rm crit}}, \quad \rho^0_{\rm crit}\equiv \frac{3H^2_0}{8\pi G},
\end{equation}
so, the Hubble parameter becomes
\begin{equation}
H^2(z)=H^2_0\left[\frac{\tilde{\zeta}}{3}+\left(\Omega^{1/2}_{m0}-\frac{\tilde{\zeta}}{3}\right)\left(1+z\right)^{3/2}\right]^2.
\label{eq:Hz}
\end{equation}

In this model the bulk viscous matter is the only component of the universe implying that the first Friedmann equation for a flat universe, $H^2=8\pi G\rho_m/3$, evaluated today gives $\Omega_{m0}=1$. With this, the Hubble parameter becomes
\begin{equation}
 H(z)=\frac{H_0}{3}\left[\tilde{\zeta}+\left(3-\tilde{\zeta}\right)\left(1+z\right)^{3/2}\right].
\label{eq:H}
\end{equation}

\subsection{Deceleration parameter $q$}

The function of the deceleration parameter $q$ is defined as
\begin{equation}
q(a)\equiv - \frac{\ddot{a}a}{\dot{a}^2} =-\frac{\ddot{a}}{a}\frac{1}{H^2}.
\label{Eq:q}
\end{equation}

With the second Friedmann equation, $\ddot{a}/a=-(4\pi G/3)\left(\rho+3p\right)$, the term $\ddot{a}/a$ can be calculated. For a matter-dominated universe with bulk viscosity, it is given by
\begin{equation}
 \frac{\ddot{a}}{a}=-\frac{4\pi G}{3}\left(\rho_m-9\zeta H \right).
\label{Eq:ddot:a}
\end{equation}

With the first Friedmann equation, $H^2=8\pi G\rho_m/3$, we have
\begin{equation}
\rho_m=\frac{3}{8\pi G}H^2(a).
\label{Eq:rho}
\end{equation}

Then, substituting Eq. \ref{Eq:rho} and the definition of $\tilde{\zeta}$ in \ref{Eq:ddot:a} we obtain
\begin{equation}
\frac{\ddot{a}}{a}=\frac{1}{2}\left(\tilde{\zeta} H_0-H(a)\right) H(a),
\label{Eq:ddot:az}
\end{equation}
and substituting \ref{Eq:ddot:az} into \ref{Eq:q}, we have
\begin{equation}
q(a, \tilde{\zeta})= \frac{1}{2} \left(1-\tilde{\zeta}\frac{H_0}{H(a)}\right).
\label{Eq:q(a)}
\end{equation}

With 
\begin{equation}
H(a, \tilde{\zeta})=\frac{\dot{a}}{a}=\frac{H_0}{3}\left(\frac{\tilde{\zeta}a^{3/2}+3-\tilde{\zeta}}{a^{3/2}}\right),
\label{Eq:H(a)}
\end{equation}
the Eq. \ref{Eq:q(a)} is given as
\begin{equation}
 q(a,\tilde{\zeta})=\frac{1}{2}\left[ \frac{3-\tilde{\zeta}(1+2a^{3/2})}{3-\tilde{\zeta}(1-a^{3/2})} \right],
\label{Eq:q(a,zeta)}
\end{equation}
then the deceleration parameter today, is given by
\begin{equation}
 q_0\equiv q(a=1,\tilde{\zeta})=\frac{1-\tilde{\zeta}}{2}.
\label{Eq:q(a=1,zeta)}
\end{equation}

Assuming the best estimated values for $\tilde{\zeta}$ from Table \ref{tabla:1} with GRBs, the resulting deceleration parameter today is $q_0=-0.4695\pm0.0324$.

We shall analyze the model presented above. Note that this is not the only proposal in this stream, but many other similar models exist in the literature, based in similar ideas \cite{Diosi:1984, Diosi:1985, Morikawa:1985, Waga:1986, Barrow:1986, Barrow:1988, Maartens:1995, Coley:1996, Brevik:2005, Feng:2009, Colistete:2007}; the conclusions we obtained then should be pertinent for models alike.

\section{Calibrating GRBs}

The main observables that can be measured when studying GRBs are its spherical equivalent energy, its peak isotropic luminosity, the peak energy of its spectrum, the photon fluence, the energy fluence, the pulse duration and the redshift of its host galaxy. Several empirical correlations among these variables can be established, see for example \cite{Reese}.

The origin, reliability and dispersion of spectral energy correlations of GRBs have been debated these late years. Some of the pros and cons we comment next on the two most used correlations, $E_{p}-E_{iso}$ (Amati) relating the rest frame energy of the spectra $E_{p}$ and the isotropic energy emitted $E_{iso}$, as well as the  $E_{\gamma}-E_{p}$ (Ghirlanda), with $E_{\gamma}=E_{iso}(1- \cos{\theta_{jet}})$ that takes into account the non-isotropic release of energy of the GRB.

Butler \textit{et al.} \cite{Butler:2009} presented a very critical analysis on the observational selection effects, as the origin of the GRBs energy correlations. They concluded that rigorously, neither  $E_{p}-E_{iso}$ nor $E_{\gamma}-E_{p}$ satisfy what should a physically-based correlation fulfil: to show a reduced scatter in the rest frame relative to the observer frame, and that must not persist if the assumed redshifts are scattered. Moreover, the discovery of some outliers to these relations has raised the suspicion that these correlations belong only to a sub-population of long GRBs or that they are an artifact of the GRBs detection process. In \cite{Shahmoradi} it is pointed out that GRBs correlations are strongly influenced by the number-biased against hard photons of the GRBs detectors.

As far as the Ghirlanda correlation $E_{\gamma}-E_{p}$, that in some analysis it turns out favored respect to the Amati's, as the tighest of the GRBs calibrations, it has the inconvenience that to be included in this relation, the GRB afterglow must have an observed jet break in its light curve, thus, only a fraction of the observed events can contribute to establishing this relation.

Amati \cite{Amati:2009} contributes to robust the $E_{p}-E_{iso}$ relation by testing two extremely energetic GRBs (GRB080916C and GRB090323) measured by \textit{Fermi}, showing that both events are fully consistent with $E_{p}-E_{iso}$ correlation. Recently, in \cite{Butler:2010} it is performed a rigorous analysis of the multivariate data, considering also selection effects of GRBs, then the authors came to the conclusion that there exists a real, intrinsic correlation between $E_{p}-E_{iso}$, but not a narrow log-log and it is strongly detector dependent.   

Having in mind the previous warnings, we consider the $E_{p}-E_{iso}$ relation along with data presented in \cite{Wei:2010} and performing the pertinent calibration, we proceed to constrain the bulk viscosity model.

Most calibrations take for granted a particular cosmological model, to remedy this circularity problem, Liang \cite{Liang:2008} proposed a cosmology-independent calibration method, consisting in calibrating at low redshifts using the SNe Ia data, mending so the few available low redshift GRBs data; the basic idea supporting the method is that light travels in the expanding universe in the same way no matter what its source be.   
The typical spectrum of the prompt emission of GRBs can be expressed as exponentially connected broken power-law, the so called Band function \cite{Band:1993}. Then we can determine spectral peak energy $E_p$, corresponding to the photon energy at maximum in $\nu F_{\nu}$ spectra. We shall apply the empirical relation  $E_p -E_{iso}$  that connects $E_p=E_{p, obs}(1+z)$ with the isotropic equivalent energy $E_{iso}$ derived by Amati \cite{Amati:2002,Amati:2008},
\begin{equation}
E_{iso}=4\pi d^2_L S_{bolo} (1+z)^{-1},
\label{Eq:Eiso}
\end{equation}
where $S_{bolo}$ is the bolometric fluence of gamma rays in the GRB at redshift $z$ and $d_L$ is the luminosity distance of the GRB (of the hosting galaxy).

Using at low redshifts the 557 Union2 SNe Ia data \cite{Union2}, Wei (2010) derived the distance moduli for the 50 low-redshift ($z<1.4$) GRBs by using a cubic interpolation from the 557 Union2 SNe Ia. Note that with the well-known relation 
\begin{equation}
 \mu= 5\log \frac{d_L}{\text{Mpc}}+25,
\label{Eq:def_mu}
\end{equation}
one can convert the distance modulus $\mu$ into luminosity distance $d_L$ (in units of Mpc). 

From Eq. \ref{Eq:Eiso} with the corresponding $S_{bolo}$ and luminosity distance $d_L$, Wei derived $E_{iso}$ for these 50 GRBs at $z<1.4$. Furthermore, with the corresponding $E_{p}$ for these 50 GRBs at $z<1.4$, he found the best fit for the Amati relation given as
\begin{equation}
\log \frac{E_{iso}}{\text{erg}}=\lambda + b\log \frac{E_{p}}{300\text{keV}},
\label{Eq:AjusteEiso}
\end{equation}
with
\begin{equation}
\lambda=52.7838\pm0.0041 \quad \text{and} \quad b=1.7828\pm0.0072.
\end{equation}

Wei's calibration was the result of an adjustment called bisector of the two ordinary least squares \cite{Isobe}. Instead of using Wei's calibration we preferred to do our own fit with a minimum least square method, using the sample given in Table 1 by Wei \cite{Wei:2010}, consisting of 50 low-redshift GRBs ($z<1.4$). We obtained the following best fit for the Amati relation,
\begin{equation}
\lambda=52.7636 \pm 0.0626, \quad b=1.6283 \pm 0.1059,
\label{Eq:NewAjusteEiso}
\end{equation}

\begin{figure*}
\centering
\includegraphics[width=6.5cm,height=6.0cm]{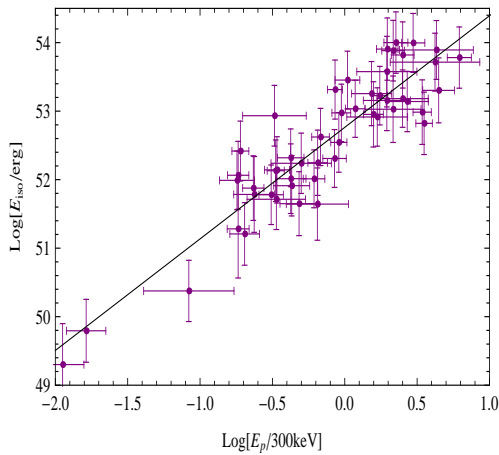}
\caption{50 GRBs at $z<1.4$ data in the $\log\left(E_p/300\text{keV}\right)-\log \left(E_{iso}/\text{erg}\right)$ plane. The best-fit calibration is the straight line with slope $b=1.6283 \pm 0.1059$. See Eq. (\ref{Eq:NewAjusteEiso}).}
\label{fig:a1}
\end{figure*}

The fit is shown in Figure \ref{fig:a1}. The errors are calculated using error propagation. Our calibration improves the one presented by Capozziello and Izzo \cite{Capo}, given by $\lambda=49.154 \pm 0.306, \quad b=1.444 \pm 0.117,$ performed with the SNe Ia Union sample of 307 objects. Moreover, being  with errors of better than one part in $10^2$, our calibration is also better than the Amati correlation presented by Liang \cite{Liang:2008}, done using the 192 SNe Ia sample. Remind that at low redshifts the calibration is obtained using SNe Ia samples, and then extrapolated to the GRBs with higher redshifts. Therefore the slight differences in calibrations are mainly the result of the different SNe Ia samples used at the low redshift calibration.

Extrapolating the calibrated Amati relation to 59 high-redshift GRBs ($z>1.4$) given in Table 2 by Wei \cite{Wei:2010}, with the corresponding $E_p$ and deriving $E_{iso}$ from the calibrated Amati relation, Eq. \ref{Eq:NewAjusteEiso} (in what follows we named MB calibration), the distance moduli $\mu$ is obtained for the extended sample of 59 GRBs at $z>1.4$ using Eqs. \ref{Eq:Eiso} and \ref{Eq:def_mu} and the respective $S_{bolo}$ also reported in \cite{Wei:2010}, in order to test the bulk viscosity model. To establish a comparison between results obtained using Wei's calibration and MB calibration, we include both in Tables \ref{tabla:1}, \ref{tabla:3}, \ref{tabla:4.1}, \ref{tabla:5}, \ref{tabla:6}. Otherwise, the GRBs analysis was done with the MB calibration.

\section{Data Analysis}

The 557 Union2 SNe Ia data compiled in \cite{Union2} and the 59 calibrated GRBs dataset in \cite{Wei:2010} are given in terms of the distance modulus $\mu_{obs}(z_i)$. The theoretical distance modulus is defined by
\begin{equation}
\mu_{th}(z;a_1,...,a_n)= 5\log \frac{d^{th}_L(z;a_1,...,a_n)}{\text{Mpc}}+25.
\end{equation}
 
On the other hand, given a parametrization $H(z;a_1,...,a_n)$ depending on $n$ parameters $a_i$, we can obtain the corresponding Hubble free luminosity distance in a flat cosmology as

\begin{equation}
d^{th}_L(z;a_1,...,a_n)=c(1+z)\int^z_0 dz' \frac{H_0}{H(z';a_1,...,a_n)}.
\label{Eq:dth}
\end{equation}

Using the maximum likelihood technique \cite{Numerical} we can find the goodness of fit for the corresponding observed $d^{obs}_L(z_i)$.
The goodness of fit corresponding to any set of parameters $a_1,...,a_n$ is determined by the probability distribution of $a_1,...,a_n$, i.e.
\begin{equation}
P(a_1,...,a_n)= \text{\textit{N}} e^{\chi^2(a_1,...,a_n)/2},
\label{Eq:Def_Probability}
\end{equation}
where
\begin{equation}
\chi^2(a_1,...,a_n)=\sum^N_{i=1}\frac{\left[\mu_{obs}(z_i)-\mu_{th}(z_i;a_1,...,a_n)\right]^2}{\sigma^2_{\mu_{obs}(z_i)}}
\label{Eq:Def_chi}
\end{equation}
and \textit{N} is a normalization factor. The parameters $\bar{a}_1,...,\bar{a}_n$ that minimize the $\chi^2$ expression \ref{Eq:Def_chi} are the most probable parameter values (the `best fit') and the corresponding $\chi^2(\bar{a}_1,...,\bar{a}_n)\equiv \chi^2_{min}$ gives an indication of the quality of the fitness for the given parametrization: the smaller $\chi^2_{min}$ is the better parametrization.
 
For the case of the 557 Union2 SNe Ia and the 59 calibrated GRBs, $\chi^2$ is given for the model of bulk viscosity as 
\begin{equation}
\chi^2_{\mu}(\tilde{\zeta}, H_0)= \sum_i \frac{\left[\mu_{obs}(z_i)-\mu_{th}(z_i, \tilde{\zeta}, H_0)\right]^2}{\sigma^2_{\mu_{obs}}(z_i)}.
\label{Eq:muSNIa}
\end{equation}

Once constructed the $\chi^2_{\mu}$ function, Eq. \ref{Eq:muSNIa}, we minimize it to find the ``best fit'' for the free parameters of the model. The probability distribution is rewritten as 
\begin{equation}
P(\tilde{\zeta}, H_0)= \text{\textit{N}} e^{\chi^2(\tilde{\zeta}, H_0)/2},
\label{Eq:Def_ProbabilityBV}
\end{equation}
where N is a normalization factor.

Besides the SNe Ia and GRBs, there are other observational data very relevant when testing cosmological models, so we consider the joint constraints from the latest observational data,
combined with the SNe Ia and GRBs, namely, the shift parameter $R$ from the WMAP 5-year and 7-year data, and the distance parameter $A$ of the measurement of the baryon acoustic oscillation (BAO) peak in the distribution of SDSS luminous red galaxies.
 
In the case of the anisotropies of the cosmic microwave background (CMB), the observational parameter that is used for the construction of the $\chi^2 $ is the shift parameter $R$. Derived in the context of the bulk viscosity model $R$ is given by
\begin{equation}
R(\tilde{\zeta})= \int^{z_*}_0 \frac{dz'}{E(z',\tilde{\zeta})}, \quad E(z,\tilde{\zeta})\equiv\frac{H(z,\tilde{\zeta})}{H_0},
\end{equation}

where the redshift of recombination is $z_*=1090.04$ from  WMAP5 data \cite{wmap5} and $z_{*}= 1091.3$ from WMAP7 data \cite{wmap7}. 
 
$\chi^2$ for $R$-CMB is defined as 
\begin{equation}
\chi^2_{CMB}(\tilde{\zeta})\equiv \left[\frac{R(\tilde{\zeta})-R_{obs}}{\sigma_{R_{obs}}}\right]^2,
\end{equation}
where $R(\tilde{\zeta})$ is the theoretical value predicted by the cosmological model and ($R_{obs}$, $\sigma_{R_{obs}}$)= (1.710,0.019) from WMAP5 \cite{wmap5} and ($R_{obs}$, $\sigma_{R_{obs}}$)= (1.725,0.018) from WMAP7 \cite{wmap7}.

In the case of baryon acoustic oscillations (BAO), it is used the acoustic peak, given by 
\begin{equation}
A(\tilde{\zeta})= \frac{1}{E^{1/3}(z_{BAO},\tilde{\zeta})}\left[\frac{1}{z_{BAO}} \int^{z_{BAO}}_0 \frac{dz'}{E(z',\tilde{\zeta})} \right]^{2/3},
\end{equation}
with $z_{BAO}=0.35$, to construct the function $\chi^2$ as
\begin{equation}
\chi^2_{BAO}(\tilde{\zeta})\equiv \left[\frac{A(\tilde{\zeta})-A_{obs}}{\sigma_{A_{obs}}}\right]^2,
\end{equation}
where and $A(\tilde{\zeta})$ is the theoretical value predicted by the cosmological model. In \cite{bao}, the value of $A$ has been determined to be $0.469(n_s/0.98)^{-0.35}\pm0.017$. Here the scalar spectral index $n_s$ is taken to be 0.963, which has been updated from WMAP7 data \cite{wmap7}. 

So, the total $\chi_T^2$ is given by
\begin{equation}
\chi_T^2\equiv\chi_T^2(\tilde{\zeta}, H_0)=\chi^2_{\mu}(\tilde{\zeta}, H_0)+\chi^2_{CMB}(\tilde{\zeta})+\chi^2_{BAO}(\tilde{\zeta}),
\end{equation}
where $\chi^2_{\mu}$ is given in Eq. \ref{Eq:muSNIa}, $\chi^2_{CMB}=(R-R_{obs})^2/\sigma^2_{R}$ and $\chi^2_{BAO}=(A-A_{obs})^2/\sigma^2_{A}$. The best-fit model parameters are determined by minimizing the total $\chi^2_T$.

\begin{figure*}
\centering
\includegraphics[width=6.5cm,height=6.0cm]{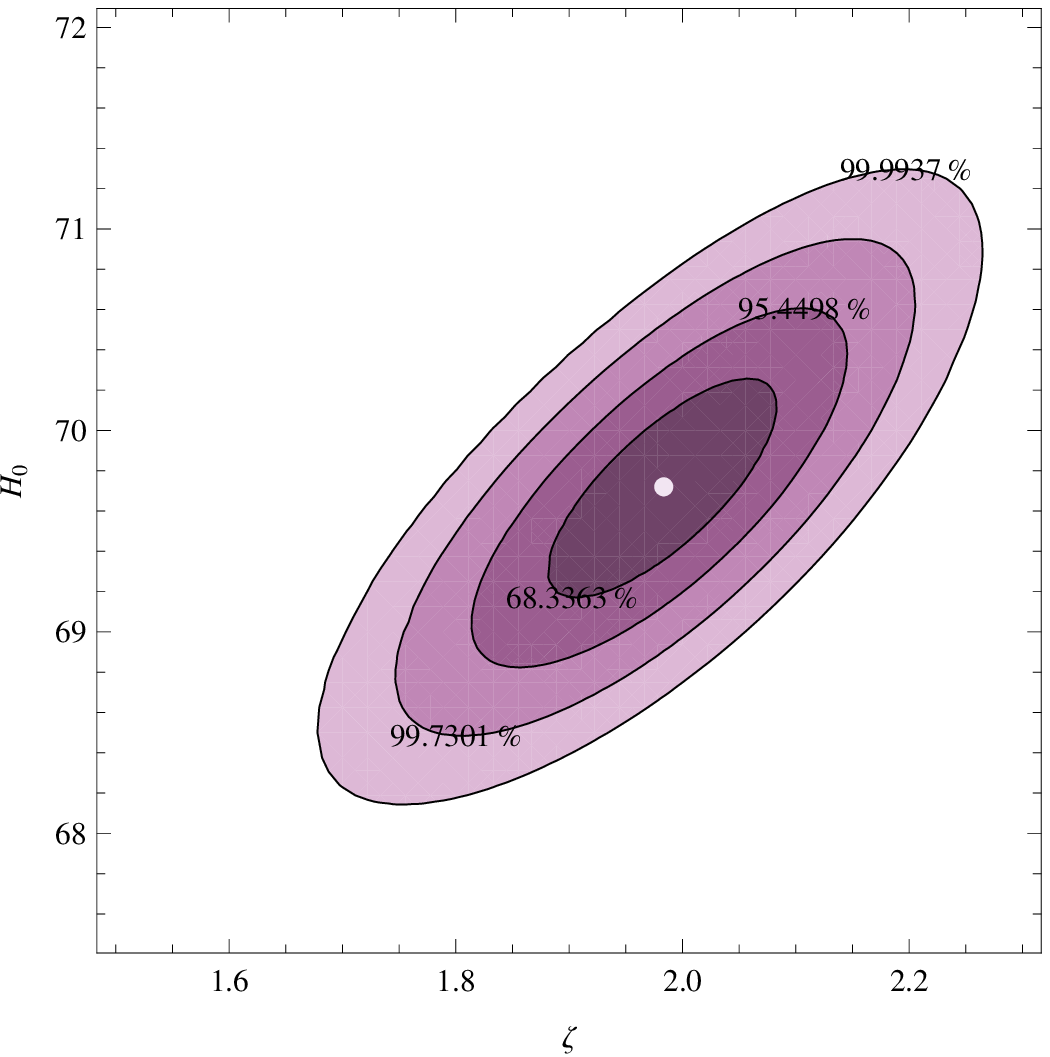}
\includegraphics[width=6.5cm,height=6.0cm]{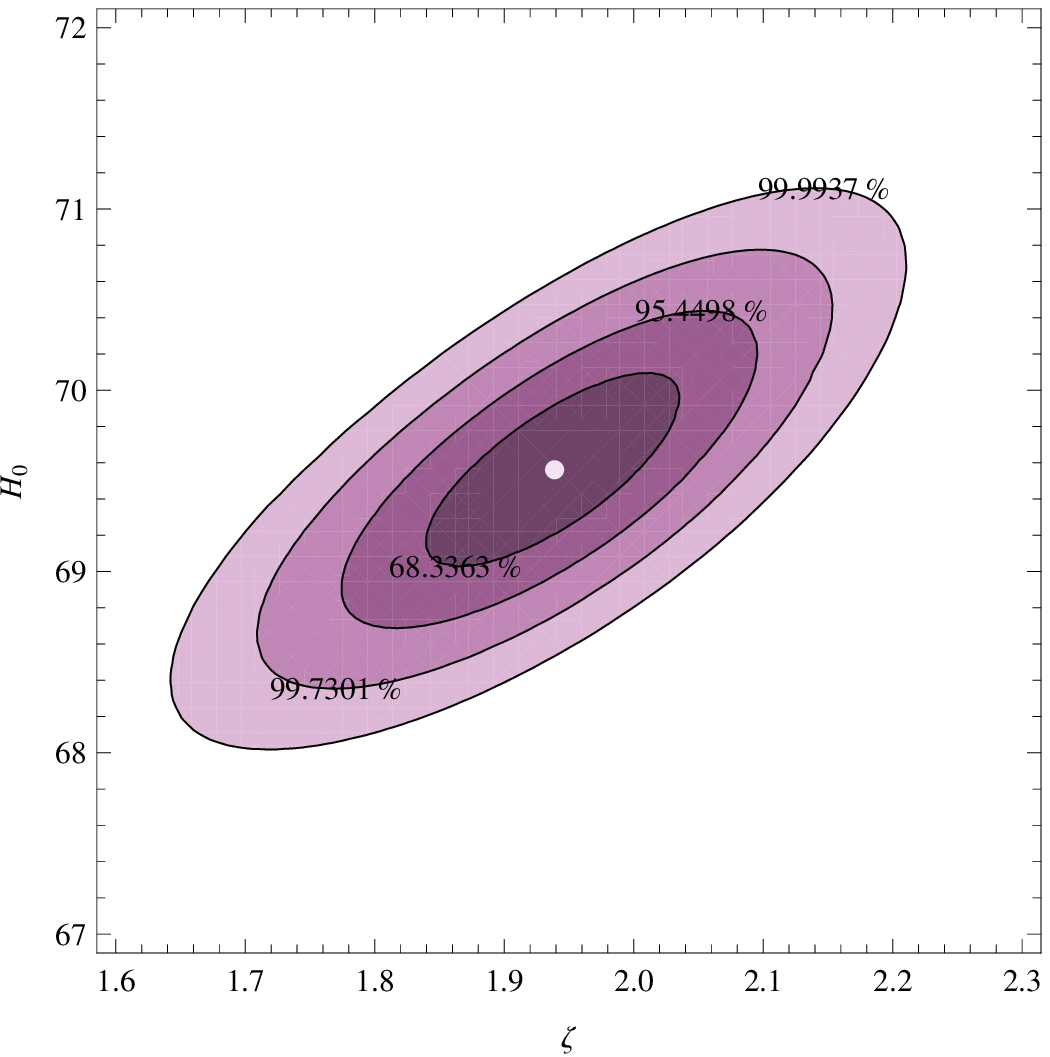}
\caption{ The joint confidence regions in the ($H_0, \tilde{\zeta}$) plane for the bulk viscosity model with $0<\tilde{\zeta}<3$. The contours correspond to $1\sigma$ - $4\sigma$ confidence regions using Union2 SNe Ia (left panel) and Union2 SNe Ia + GRBs (right panel), respectively. GRBs have been calibrated using the MB Calibration. The best estimated values and confidence intervals using the Union2 SNe Ia data set are $\tilde{\zeta}= 1.9835 \pm 0.0668$ and $H_0=  69.7130 \pm 0.3572$ and  using  Union2 SNe Ia + GRBs data set are $\tilde{\zeta}=1.9389 \pm 0.0647$ and $H_0=69.5616\pm 0.3523$, which are pointed with a dot.}
\label{fig:1}
\end{figure*}

\begin{table*}
\centering
\begin{tabular}{|c|c|c|c|}\hline
\multirow{2}{*}{} & \multirow{2}{*}{SNe Ia} & \multicolumn{2}{|c|}{SNe Ia + GRBs} \\ \cline{3-4}
 & & Wei's Calibration & MB Calibration \\ \hline
 $\tilde{\zeta}$ & 1.9835 $\pm$ 0.0668 & 1.9684 $\pm$ 0.0654&1.9389$\pm$0.0647 \\
$H_0$           & 69.7130 $\pm$ 0.3572& 69.6610$\pm$ 0.3552&69.5616$\pm$0.3523\\
$z_t$           & 1.4788 $\pm$ 0.1679 & 1.4421 $\pm$ 0.1573&1.3726$\pm$0.1492\\
$q_0$           & -0.4918 $\pm$ 0.0334& -0.4842$\pm$ 0.0327&-0.4695$\pm$0.0324\\
$\chi^2_{min}$  &544.5880             & 570.1470&586.7870\\
$\chi^2_{d.o.f.}$& 0.9830              & 0.9301&0.9572\\ \hline
\end{tabular}
\caption{The best-fit value for the bulk viscosity model parameters ($H_0$, $\tilde{\zeta}$), without priors, with 1$-\sigma$ uncertainties, $\chi_{min}^2$, $\chi^2_{d.o.f}$, as well as $z_t$ and $q_0$ using SNe Ia, and SNe Ia + GRBs, where GRBs have been calibrated using the Wei's Calibration or the Montiel-Bret\'on (MB) Calibration. $H_0$ is in units of km s$^{-1}$ Mpc$^{-1}$ and $\tilde{\zeta}$, $z_t$ and $q_0$ are dimensionless. The subscript ``t'' stands for ``transition'' and the subscript ``d.o.f''. stands for ``\textit{degrees of freedom}''. The confidence intervals are shown in Figure \ref{fig:1}.}
\label{tabla:1}
\end{table*}

\begin{table*}
\centering 
\begin{tabular}{|c|c|c|c|c|}\hline

               &\tiny{SNe Ia + CMB5 + BAO}&\tiny{SNe Ia + CMB7 + BAO}& \tiny{SNe Ia + CMB5 + BAO + GRBs}   &\tiny{SNe Ia + CMB7 + BAO + GRBs}\\ \hline
$\tilde{\zeta}$ &-0.06523$\pm$0.03552 &-0.07420$\pm$0.03423 &-0.05898$\pm$0.03532 & -0.06837$\pm$0.03405 \\
$H_0$           &62.89780$\pm$0.23522 &62.87290$\pm$0.23347 &62.88625$\pm$0.23501 & 62.86002$\pm$0.16042 \\
$z_t$           &-0.87809$\pm$0.04331 &-0.86742$\pm$0.03979 &-0.88586$\pm$0.04469 & -0.87430$\pm$0.04080 \\
$q_0$           &0.53262$\pm$0.01776&0.53710$\pm$0.01712 &0.52949$\pm$0.01766 & 0.53419$\pm$0.01703 \\
$\chi^2_{min}$  &1279.08 &1273.84 &1323.06&1317.91 \\
$\chi^2_{d.o.f}$&2.3005 &2.2911 &2.1513&2.1429 \\ \hline
\end{tabular}
\caption{The best-fit value for the bulk viscosity model parameters ($H_0$, $\tilde{\zeta}$), without priors, with 1$-\sigma$ uncertainties, $\chi_{min}^2$, $\chi^2_{d.o.f}$, as well as $z_t$ and $q_0$ using SNe Ia + CMB5 + BAO, SNe Ia + CMB7 + BAO, SNe Ia + CMB5 + BAO + GRBs and SNe Ia + CMB7 + BAO + GRBs, respectively. CMB5 and CMB7 stand for CMB data from 5-years and 7-years from WMAP.  }
\label{tabla:2}
\end{table*}

\subsection{Marginalization over $H_0$}

In the statistical process, if some parameters are known, this information can be used to `marginalize' the known parameters, i.e. average the probability distribution \ref{Eq:Def_Probability} around the known value with an appropriate `prior' probability distribution. In this work, we marginalize over the Hubble constant $H_0$ in order to have $\tilde{\zeta}$ as the only free parameter of the model. We use two different priors to marginalize $H_0$: constant and Dirac Delta priors.
\begin{enumerate}
 \item \textit{Constant prior} over $H_0$.

In this case, we should assume that $H_0$ does not have any preferred value a priori, i.e, it has a constant prior probability distribution function. Instead of minimizing the function $\chi^2$ given by Eq. \ref{Eq:muSNIa}, we minimize $\chi_{cp}^2$ given as 
\begin{equation}
\chi^2_{cp}(\tilde{\zeta})\equiv A(\tilde{\zeta})-\frac{\left[B(\tilde{\zeta})+\ln(10)/5\right]^2}{C},
\end{equation}
 where 
\begin{eqnarray}
&A\equiv \sum^n_{i=1}\left(\frac{\tilde{\mu}^{th}_i-\mu^{obs}_i}{\sigma_i}\right)^2, \nonumber \\
& B\equiv \sum^n_{i=1}\frac{\tilde{\mu}^{th}_i-\mu^{obs}_i}{\sigma^2_i},     \nonumber \\
& C\equiv\sum^n_{i=1}\frac{1}{\sigma^2_i},
\end{eqnarray}

with
\begin{equation}
\tilde{\mu}^{th}\equiv5\log \left[ (1+z)\int^z_0 \frac{dz'}{E(z', \tilde{\zeta})}\right]+25.
\end{equation}

This new $\chi^2_{cp}$ function does not depend on $H_0$ anymore, the label ``cp'' stands for \textit{constant prior} for $H_0$. For more detailed analyzes see the Appendix A of \cite{Nucamendi:2009}.

\begin{table*}
\centering
\begin{tabular}{|c|c|c|c|}\hline
\multirow{2}{*}{} & \multirow{2}{*}{SNe Ia} & \multicolumn{2}{|c|}{SNe Ia + GRBs} \\ \cline{3-4}
 & & Wei's Calibration & MB Calibration \\ \hline
$\tilde{\zeta}$ & 1.9838 $\pm$ 0.0668& 1.9686 $\pm$ 0.0655& 1.9844$\pm$0.0666\\
$z_t$           & 1.4795 $\pm$ 0.1642& 1.4425 $\pm$ 0.1576&1.4810$\pm$0.1640\\
$q_0$           & -0.4919$\pm$ 0.0334& -0.4843$\pm$ 0.0328&-0.4922$\pm$0.0333\\
$\chi^2_{min}$  & 561.3220           & 586.8830 &613.6350\\
$\chi^2_{d.o.f}$& 1.0114             & 0.9558 &0.9994\\ \hline
\end{tabular}
\caption{The best-fit value of $\tilde{\zeta}$ with 1-$\sigma$ uncertainties, and $\chi^2_{min}$, $\chi^2_{d.o.f.}$, as well as $z_t$ and $q_0$ for the model of bulk viscosity assuming a constant prior over $H_0$. These results were obtained using SNe Ia and SNe Ia + GRBs, respectively.}
\label{tabla:3}
\end{table*}

\begin{table*}
\centering
\begin{tabular}{|c|c|c|c|c|} \hline
                &\tiny{SNe Ia + CMB5 + BAO}&\tiny{SNe Ia + CMB7 + BAO}&\tiny{SNe Ia + CMB5 + BAO + GRBs} &\tiny{SNe Ia + CMB7 + BAO + GRBs}\\ \hline
$\tilde{\zeta}$ &-0.06518$\pm$0.03552 & -0.07415$\pm$0.03423 & -0.05893$\pm$0.03532 &-0.06833$\pm$0.03406\\
$z_t$           &-0.87816$\pm$0.04332 & -0.86748$\pm$0.03979 &-0.88592$\pm$0.04470 &-0.87435$\pm$0.04082\\
$q_0$           &0.53259$\pm$0.01776  & 0.53708$\pm$0.01712  &0.52947$\pm$0.01766 &0.53417$\pm$0.01703\\
$\chi^2_{min}$  &1296.02            & 1290.78            &1340.00           & 1334.85\\
$\chi^2_{d.o.f}$& 2.3268             & 2.3174            &2.1753            & 2.1670\\ \hline
\end{tabular}
\caption{The best-fit value of $\tilde{\zeta}$ with 1-$\sigma$ uncertainties, and $\chi^2_{min}$, $\chi^2_{d.o.f.}$, as well as $z_t$ and $q_0$ for the model of bulk viscosity assuming a constant prior over $H_0$ with SNe Ia + CMB5 + BAO, SNe Ia + CMB7 + BAO, SNe Ia + CMB5 + BAO + GRBs and SNe Ia + CMB7 + BAO + GRBs, respectively.}
\label{tabla:4}
\end{table*}

\begin{table*}
\centering
\begin{tabular}{|c|c|c|c|}\hline
\multirow{2}{*}{} & \multirow{2}{*}{SNe Ia + BAO} & \multicolumn{2}{|c|}{SNe Ia + BAO + GRBs} \\ \cline{3-4}
 & & Wei's Calibration & MB Calibration \\ \hline
$\tilde{\zeta}$ & 1.8693 $\pm$ 0.0625& 1.8606 $\pm$ 0.0616&1.8388$\pm$0.0654\\
$z_t$           & 1.2194 $\pm$ 0.1313& 1.2013 $\pm$ 0.1279&1.1566$\pm$0.1321\\
$q_0$           & -0.4347$\pm$ 0.0313& -0.4303$\pm$ 0.0308&-0.4194$\pm$0.0327\\
$\chi^2_{min}$  & 578.138            & 603.065&618.3780 \\
$\chi^2_{d.o.f}$& 1.0398             & 0.9806&1.0055\\ \hline
\end{tabular}
\caption{The best-fit value of $\tilde{\zeta}$ with 1-$\sigma$ uncertainties, and $\chi^2_{min}$, $\chi^2_{d.o.f.}$, as well as $z_t$ and $q_0$ for the model of bulk viscosity assuming a constant prior over $H_0$ using SNe Ia + BAO and SNe Ia + BAO + GRBs, respectively.}
\label{tabla:4.1}
\end{table*}

 \item \textit{ Dirac Delta prior} over $H_0$.

We assume that $H_0$ has a specific value (suggested by some other independent observation). In this case, the probability distribution has the form of a Dirac delta centered at that specific value. In particular, we choose $H_0=70.5$ km s$^{-1}$ Mpc$^{-1}$, as reported by 5 year WMAP data, and $H_0=73.8\pm2.4$km s$^{-1}$ Mpc$^{-1}$ as suggested by the observations of the Hubble Space Telescope (HST) \cite{HST}.

Using a prior with the form of a Dirac delta for $H_0$  
\begin{equation}
 P(H_0)=\delta(H_0-H^*_0),
\end{equation}
the expression \ref{Eq:Def_ProbabilityBV} becomes
\begin{equation}
P(\tilde{\zeta})= \text{\textit{N}}~\cdotp e^{-\chi^2(\tilde{\zeta},H^*_0)/2},
\end{equation}
where \textit{N} is a normalization constant. 

\end{enumerate}

\begin{table*}
\centering
\begin{tabular}{|c|c|c|c|} \hline
\multirow{2}{*}{} & \multirow{2}{*}{SNe Ia} & \multicolumn{2}{|c|}{SNe Ia + GRBs} \\ \cline{3-4}
 & & Wei's Calibration & MB Calibration \\ \hline
$\tilde{\zeta}$ & 2.0910$\pm$0.0429  & 2.0807 $\pm$ 0.0423&2.0630$\pm$0.0418 \\
$z_t$           & 1.7662$\pm$0.1249  & 1.7365 $\pm$ 0.1210&1.6865$\pm$0.1162\\
$q_0$           & -0.5455$\pm$0.0215 &-0.5404$\pm$ 0.0212&-0.5315$\pm$0.0212\\
$\chi^2_{min}$  & 549.381            & 575.706     & 593.857     \\
$\chi^2_{d.o.f}$& 0.9899             & 0.9376&0.9672 \\ \hline
\end{tabular}
\caption{The best-fit value of $\tilde{\zeta}$ with 1-$\sigma$ uncertainties, and $\chi^2_{min}$, $\chi^2_{d.o.f.}$, as well as $z_t$ and $q_0$ were estimated using SNe Ia and SNe Ia + GRBs data set, respectively,  for the model of bulk viscosity assuming a Dirac delta prior over $H_0$ located at $H_0=70.5$ km s$^{-1}$ Mpc$^{-1}$, as reported by 5 year WMAP data.}
\label{tabla:5}
\end{table*}

\begin{table*}
\centering
\begin{tabular}{|c|c|c|c|} \hline
\multirow{2}{*}{} & \multirow{2}{*}{SNe Ia} & \multicolumn{2}{|c|}{SNe Ia + GRBs} \\ \cline{3-4}
 & & Wei's Calibration & MB Calibration \\ \hline
$\tilde{\zeta}$ & 2.4909$\pm$0.0383  & 2.4696 $\pm$ 0.0375& 2.4440$\pm$0.0370\\
$z_t$           & 3.5750$\pm$0.2764  & 3.4262 $\pm$ 0.2534 & 3.2596$\pm$0.2320\\
$q_0$           &-0.7455$\pm$0.0192  &-0.7348$\pm$ 0.0188& -0.7220$\pm$0.0185\\
$\chi^2_{min}$  & 671.682            & 703.588     & 729.296     \\
$\chi^2_{d.o.f}$& 1.2102             & 1.1459 & 1.1878\\ \hline
\end{tabular}
\caption{The best-fit value of $\tilde{\zeta}$ with 1-$\sigma$ uncertainties, and $\chi^2_{min}$, $\chi^2_{d.o.f.}$, as well as $z_t$ and $q_0$ were estimated using SNe Ia and SNe Ia + GRBs data set, respectively,  for the model of bulk viscosity assuming a Dirac delta prior over $H_0$ located at $H_0=73.8$ km s$^{-1}$ Mpc$^{-1}$, as suggested by the observations of the Hubble Space Telescope (HST).}
\label{tabla:6}
\end{table*}

\begin{table*}
\centering
\begin{tabular}{|c|c|c|c|c|} \hline
                &\tiny{SNe Ia+CMB5+BAO}&\tiny{SNe Ia+CMB7+BAO} & \tiny{SNe Ia + CMB5 + BAO + GRBs} & \tiny{SNe Ia + CMB7 + BAO + GRBs}\\ \hline
$\tilde{\zeta}$ & 2.99393$\pm$0.00020& 2.99402$\pm$0.00011   &2.99393$\pm$0.00013    & 2.99402$\pm$0.00011 \\
$z_t$           & 98.09581$\pm$2.18115& 99.08961$\pm$1.22986  &98.09581$\pm$1.41775    & 99.08961$\pm$1.22986\\
$q_0$           & -0.99697$\pm$0.00010 & -0.99701$\pm$0.00006   &-0.99697$\pm$0.00007  & -0.99701$\pm$0.00006\\
$\chi^2_{min}$  & 1883.15     & 1883.27             & 2029.31            & 2029.47\\
$\chi^2_{d.o.f}$& 3.3809      & 3.3811               & 3.2943             & 3.2946\\ \hline
\end{tabular}
\caption{The best-fit value of $\tilde{\zeta}$ with 1-$\sigma$ uncertainties, and $\chi^2_{min}$, $\chi^2_{d.o.f.}$, as well as $z_t$ and $q_0$ were estimated using SNe Ia + CMB5 + BAO, SNe Ia + CMB7 + BAO, SNe Ia + CMB5 + BAO + GRBs and SNe Ia + CMB7 + BAO + GRBs, respectively,  for the model of bulk viscosity assuming a Dirac delta prior over $H_0$ located at $H_0=70.5$ km s$^{-1}$ Mpc$^{-1}$, as reported by 5 year WMAP data.}
\label{tabla:7}
\end{table*}

\begin{table*}
\centering
\begin{tabular}{|c|c|c|c|c|} \hline
                &\tiny{SNe Ia+CMB5+BAO}&\tiny{SNe Ia+CMB7+BAO} & \tiny{SNe Ia + CMB5 + BAO + GRBs} & \tiny{SNe Ia + CMB7 + BAO + GRBs}\\ \hline
$\tilde{\zeta}$ &2.99439$\pm$0.00010& 2.99447$\pm$0.00010   &2.99439$\pm$0.00005&2.99447$\pm$0.00011  \\
$z_t$           &103.45206$\pm$1.24359& 104.45889$\pm$1.27371&103.45206$\pm$0.62179&104.45889$\pm$1.40107 \\
$q_0$           &-0.99720$\pm$0.00005& -0.99724$\pm$0.00005   &-0.99720$\pm$0.00003&-0.99724$\pm$0.00005 \\
$\chi^2_{min}$  & 1641.81        & 1641.88             &  1775.06         &1775.16 \\
$\chi^2_{d.o.f}$& 2.9476         & 2.9477              &  2.8816           &2.8818 \\ \hline
\end{tabular}
\caption{The best-fit value of $\tilde{\zeta}$ with 1-$\sigma$ uncertainties, and $\chi^2_{min}$, $\chi^2_{d.o.f.}$, as well as $z_t$ and $q_0$ were estimated using SNe Ia + CMB5 + BAO, SNe Ia + CMB7 + BAO, SNe Ia + CMB5 + BAO + GRBs and SNe Ia + CMB7 + BAO + GRBs, respectively,  for the model of bulk viscosity assuming a Dirac delta prior over $H_0$ located at $H_0=73.8$ km s$^{-1}$ Mpc$^{-1}$, as suggested by the observations of the HST.}
\label{tabla:8}
\end{table*}
\section{Discussion}
In the previous section, we have obtained the constraints of the bulk viscosity model with the latest observational data: the combination of 557 Union2 SNe Ia data set in the range $0.02<z<1.4$ \cite{Union2}, 59 calibrated GRBs data set at higher redshifts, the shift parameter $R$ from the WMAP5 or WMAP7 data \cite{wmap5, wmap7}, and the distance parameter $A$ of SDSS luminous red galaxies \cite{bao}.

As we mentioned above, the best-fit model parameters are determined by minimizing the total $\chi^2_T$. For comparison, SNe Ia and SNe Ia + CMB + BAO without GRBs have been used to show which is the contribution of GRBs to the joint cosmological  constraints. In addition, some different data sets such as SNe Ia + BAO and SNe Ia + BAO + GRBs have also been jointly considered to inspect if the BAO or CMB data are responsible of spoiling the statistics. 

In Figure \ref{fig:1}, we show the joint confidence regions in the ($\tilde{\zeta}$, $H_0$) plane for the bulk viscosity model with $\tilde{\zeta}=$cte. Using 557  SNe Ia + 59 GRBs (with high redshifts), the 1-$\sigma$ confidence region for  ($\tilde{\zeta}$, $H_0$) of the bulk viscosity model is ($\tilde{\zeta}$, $H_0$) = (1.9389$\pm$0.0647, 69.5616$\pm$0.3523 ) with $\chi^2_{d.o.f.}=$0.9572. For comparison, fitting results from the data without GRBs are also given in Figure \ref{fig:1} (to the left). With 557 SNe Ia, the best-fit values are ($\tilde{\zeta}$, $H_0$) = (1.9835$\pm$0.0668, 69.7130$\pm$0.3572) and $\chi^2_{d.o.f.}=$0.9830. We present in Table \ref{tabla:1} the best-fit value of $\tilde{\zeta}$ and $H_0$ with 1-$\sigma$ uncertainties, $\chi^2_{min}$ and $\chi^2_{d.o.f.}$, as well as, the deceleration parameter $q_0$ and the redshift, $z_t$, of the transition between the decelerated-accelerated expansion epochs, as functions of $\tilde{\zeta}$; the deceleration parameter is according to the $\Lambda$CDM prediction. Moreover, from these values for $\tilde{\zeta}$ and $H_0$, we see that no significative difference arise from probing with SNe Ia or with SNe Ia + GRBs. Then, the validity of the constant bulk viscosity model is extended to redshifts of $z\sim8.1$.

In Table \ref{tabla:2}, we present the best-fit value of $\tilde{\zeta}$ and $H_0$ with 1-$\sigma$ uncertainties, and $\chi^2_{min}$, $\chi^2_{d.o.f.}$, as well as, $q_0$ and $z_t$ but this time using the combination SNe Ia + CMB5 + BAO, SNe Ia + CMB7 + BAO, SNe Ia + CMB5 + BAO + GRBs and SNe Ia + CMB7 + BAO + GRBs, respectively. In all the cases, we found $\tilde{\zeta}<0$ that is a value unacceptable since it is associated to violation of the second law of thermodynamics, however the inclusion of GRBs significantly improves the statistics. In Table \ref{tabla:2}, check first and third column and second and fourth, including GRBs we obtained a better value in $\chi^2_{d.o.f.}$ The negative values of $\tilde{\zeta}$ has as consequence that $q_0>0$ (see Eq. \ref{Eq:q(a=1,zeta)}) and $z_t= \left( 2\tilde{\zeta}/(3-\tilde{\zeta})\right)^{2/3}-1$ becomes negative.
Moreover, being $\tilde{\zeta}<0$ the pressure turns out to be positive since $P^*=-3\zeta H$. The meaning of this is that we can not extrapolate the viscous fluid behaviour to so far epochs.

We can conclude that the model has limited applicability when extrapolated to very large redshifts like that of the recombination epoch, i.e. when the statistical tests include the CMB. Notice that all the parameter suffer a drastic change in their estimated values, the unreliability of them is measured by the increasing in $\chi^2_{d.o.f.}$ and $\chi^2_{min}$ as 2 and as doubled data, respectively. However, the statistics is better when included GRBs than just SNe Ia, i.e. SNe Ia + CMB + BAO is worst than SNe Ia + CMB + BAO + GRBs. 

In the following Tables \ref{tabla:3}, \ref{tabla:4} and \ref{tabla:4.1} are presented the best-fit values of $\tilde{\zeta}$ with 1-$\sigma$ uncertainties, and $\chi^2_{min}$, $\chi^2_{d.o.f.}$, as well as $z_t$ and $q_0$ for the model of bulk viscosity assuming a constant prior over $H_0$. In Table \ref{tabla:3} are shown the results obtained using SNe Ia and SNe Ia + GRBs, which are very similar to those obtained without prior over $H_0$. The effect of assuming a constant prior on $H_0$, with and without GRBs, is a best value of $\chi^2_ {d.o.f.}$ In Table \ref{tabla:4} the results are using SNe Ia + CMB5 + BAO, SNe Ia + CMB7 + BAO, SNe Ia + CMB5 + BAO + GRBs and SNe Ia + CMB7 + BAO + GRBs. The obtained values are almost the same with or without assuming a constant prior on $H_0$ but the best-fit is obtained without prior, and statistics is better when GRBs are included. 

Until now we have seen that when we consider only SNe Ia + GRBs we obtain $\tilde{\zeta}\geq 0$ and a reasonable value for the Hubble constant $H_0$. However, when the same analysis is  performed using SNe Ia, CMB5 or CMB7, and BAO with or without GRBs , we obtain negative values of $\tilde{\zeta}$ that disagree with the second law of thermodynamics and moreover the estimation for $q_0$ and $z_t$ is not good and also $\chi^2_{d.o.f.}$ gets worst. 

Since the redshift of CMB corresponds to earlier epochs of the universe while BAO refers to more recent ones ($z \sim 0.35$), our guess is that it is the inclusion of CMB that ruins our statistics. The aim of Table \ref{tabla:4.1} is to prove it. We derived the constraints for the model of bulk viscosity using SNe Ia + BAO and  SNe Ia + BAO + GRBs and we found, in both cases, that the value of $\tilde{\zeta}$ is in perfect agreement with the value predicted by the model to have an accelerated expansion epoch, that is to say, $1<\tilde{\zeta}<3$, and also with this value of $\tilde{\zeta}$ we have an accelerated universe in the present with a transition between decelerated-accelerated expansion at $z\approx1.2$. This result agrees with the analysis on perturbation dynamics and the calculated matter power spectrum done by Zimdahl \cite{Zimdahl:2009} on the bulk viscosity model, that turned out compatible with the data from 2dFGRS and SDSS surveys.

Finally, we analyze and constraint the viability of the model assuming a Dirac delta prior over $H_0$ located at $H_0=$70.5 km s$^{-1}$ Mpc$^{-1}$ as reported by 5-year WMAP data and at $H_0=$73.8 km s$^{-1}$ Mpc$^{-1}$ as suggested by the observations of the HST, the results are shown from Table \ref{tabla:5} to Table \ref{tabla:8}. 

Again, the results obtained with $H_0=$70.5 km s$^{-1}$ Mpc$^{-1}$ using SNe Ia and using the joint SNe Ia + GRBs, are in agreement with the model and good values are obtained for $z_t$ and $q_0$. However when we use $H_0=$73.8 km s$^{-1}$ Mpc$^{-1}$ we obtained values for $\tilde{\zeta}$ in the expected range but with a slight increment in the value of $\chi^2_{d.o.f.}$ This is shown in Table \ref{tabla:5} and \ref{tabla:6}. On the other hand, when we used SNe Ia, CMB data (WMAP5 or WMAP7) and BAO with or without GRBs, assuming a Dirac delta prior over $H_0$ located at $H_0=$70.5 km s$^{-1}$ Mpc$^{-1}$ or $H_0=$73.8 km s$^{-1}$ Mpc$^{-1}$, we found values for $\tilde{\zeta}$ in the range suggested by the model but with bad statistic, which is reflected in the value of $\chi^2_{d.o.f.}$ Moreover, since $\tilde{\zeta}\sim 3$, then $z_t= \left( 2\tilde{\zeta}/(3-\tilde{\zeta})\right)^{2/3}-1$ turns out to be enormous, and $q_0\sim -1$, for these reasons we conclude that the statistics for a Delta Dirac prior on $H_0$ is not reliable. It should be noted that when we use SNe Ia + CMB5 + BAO + GRBs  or SNe Ia + CMB7 + BAO + GRBs, i.e. when GRBs are included, significantly improve the value of $\chi^2_{d.o.f.}$, but anyway, we have a worse fit to data each time that CMB is considered. See Tables \ref{tabla:7} and \ref{tabla:8}.

Lastly, if we compare the results obtained using Wei's Calibration or MB Calibration for GRBs, as shown in Tables \ref{tabla:1}, \ref{tabla:3}, \ref{tabla:4.1}, \ref{tabla:5} and \ref{tabla:6}, we notice that in most cases $\chi_{d.o.f.}^2$ is closer to 1 when using MB calibration, that is to say, the latest calibration provide a better fit. 

\section{Conclusions}

Modeling the dark sector of the universe with a bulk viscous fluid with 
pressure
$p=- \tilde{\zeta} \theta$, with a constant bulk viscosity $\tilde \zeta$, 
is tested with
several cosmological data: SNe Ia, CMB and BAO jointly with the calibrated 
sample of GRBs.
The model is in good agreement with observational data from SN Ia, BAO and 
GRBs, up to the range of $z=8.1$ with a chi-square statistics of 
$\chi^2=1.01$.

Probing with SNe Ia and GRBs we obtained as the present Hubble parameter 
$H_0=69.56\text{km}\text{s}^{-1}\text{Mpc}^{-1}$, the deceleration parameter $q_0=-0.47$, both in good agreement with the accepted values; as the redshift transition parameter we obtain 
$z_t=1.37$. When the baryon acoustic oscillations (BAO) data are 
included in the data-model confrontation, we obtain a viscosity parameter 
$\tilde \zeta=1.84$, included in the correct range, and improved values 
for $z_t=1.16$ and $\chi^2=1.01$. The results with BAO confirm the analysis 
performed by Hip\'olito-Ricaldi (2009) \cite{Zimdahl:2009} on the matter power spectrum 
showing the compatibility of the bulk viscosity models with the 2dFGRS and 
SDSS data.

However the model does not pass the CMB test, exhibiting
$\chi^2=2.15$ for CMB5 and $\chi^2=2.14$ for CMB7, besides a negative bulk 
viscosity constant,
violating so the second law of thermodynamics. These results are 
consequence of neglecting
a radiative component in the universe, that in so far epochs as the 
recombination, plays an important role. Thus, when testing so far epochs a 
radiation component should be added to the universe content,
a component that is conserved independently of the rest of matter 
considered.

To conclude, constant bulk viscosity matter reproduce observational SN Ia, 
GRBs and BAO data in good agreement ($\chi^2=1.01$), being then a reliable 
model not only for the recent state of the universe, but for epochs where 
structure already existed ($z=8.1$). It is an open questions if the model 
could account for the CMB data if a radiation component is included in the 
considered matter content of the universe.

\begin{acknowledgments}
A.M. acknowledges financial support from Conacyt-M\'exico, through a PhD. grant.
\end{acknowledgments}

\end{document}